\documentclass{article} 
\usepackage{iclr2026_conference,times}

\usepackage{amsmath,amsfonts,bm}

\def\eqref#1{equation~\ref{#1}}

\def\1{\bm{1}}

\DeclareMathAlphabet{\mathsfit}{\encodingdefault}{\sfdefault}{m}{sl}
\SetMathAlphabet{\mathsfit}{bold}{\encodingdefault}{\sfdefault}{bx}{n}

\usepackage{hyperref}
\usepackage{url}
\usepackage[utf8]{inputenc} 
\usepackage[T1]{fontenc}    
\usepackage{hyperref}       
\usepackage{booktabs}       
\usepackage{amsmath}
\usepackage{amsfonts}       
\usepackage{nicefrac}       
\usepackage{microtype}      
\usepackage{xcolor}         
\usepackage[capitalize,noabbrev]{cleveref}
\usepackage{float}
\usepackage{algorithm,algpseudocode}
\usepackage{placeins}
\usepackage{graphicx}
\usepackage{multirow}
\usepackage{multicol}
\usepackage{enumitem}

\usepackage{authblk}

\algrenewcommand\algorithmicrequire{\textbf{Input:}}
\algrenewcommand\algorithmicensure{\textbf{Output:}}

\title{Privacy-Preserving Mechanisms Enable Cheap Verifiable Inference of LLMs}

\author[ ]{Arka Pal\textsuperscript{\normalfont{1}}\thanks{Correspondence to: \texttt{arka@ritual.net}}}
\author[1]{Louai Zahran}
\author[1,2]{William Gvozdjak}
\author[1]{Akilesh Potti}
\author[1,3]{Micah Goldblum}

\affil[1]{Ritual}
\affil[2]{MIT}
\affil[3]{Columbia University}

\iclrfinalcopy 
\begin{document}

\maketitle

\begin{abstract}
As large language models (LLMs) continue to grow in size, fewer users are able to host and run models locally. This has led to increased use of third-party hosting services. However, in this setting, there is a lack of guarantees on the computation performed by the inference provider. For example, a dishonest provider may replace an expensive large model with a cheaper-to-run weaker model and return the results from the weaker model to the user. Existing tools to verify inference typically rely on methods from cryptography such as zero-knowledge proofs (ZKPs), but these add significant computational overhead, and remain infeasible for use for large models. In this work, we develop a new insight -- that given a method for performing \emph{private} LLM inference, one can obtain forms of \emph{verified} inference at marginal extra cost. Specifically, we propose two new protocols which leverage privacy-preserving LLM inference in order to provide guarantees over the inference that was carried out. Our approaches are cheap, requiring the addition of a few extra tokens of computation, and have little to no downstream impact. As the fastest privacy-preserving inference methods are typically faster than ZK methods, the proposed protocols also improve verification runtime. Our work provides novel insights into the connections between privacy and verifiability in LLM inference. We open-source our code at \href{https://github.com/louai-ritual/priveri}{https://github.com/louai-ritual/priveri}.

\end{abstract}

\section{Introduction}

Large language models (LLMs) have increased significantly in size over the last few years. Recent models achieving cutting-edge performance \citep{deepseekai2025deepseekr1incentivizingreasoningcapability, qwen2025qwen25technicalreport, kimiteam2025kimik2openagentic}, for example, now often contain hundreds of billions of parameters. The hardware requirements to run these models are often too high for individuals, or even organizations, to run on their own, leading to a significant growth in demand for third-party LLM inference providers. However, this trend raises critical concerns about the integrity and trustworthiness of the services provided, particularly in the growing decentralized inference space. In this setting, any entity with surplus computational resources can offer to complete computational tasks, such as LLM inference, for another user. As the providers in this setting are often not subject to strict vetting, it is imperative to ensure that the service paid for is actually one that is performed by the provider.

Traditionally, the verification of outsourced computation has been addressed through cryptographic methods, such as zero-knowledge proofs (ZKPs). Although offering strong theoretical guarantees, these methods often introduce substantial computational overhead for either the prover (the inference provider) or the verifier (the user), or both. Despite significant progress in recent years, the state-of-the-art for ZK verification of LLM inference remains thousands of times slower than vanilla inference \citep{sun2024zkllmzeroknowledgeproofs}, rendering it infeasible for large models, which are particularly likely to be in demand for third-party inference provision.

A related concern for third-party compute provision is that of \emph{privacy-preservation}. Performing LLM inference for another party requires the user to share their prompts, resulting in a loss of privacy. Therefore, a seemingly orthogonal line of work in recent years has focused on privacy-preserving computation. These include methods such as secure multi-party computation (SMPC) and fully homomorphic encryption (FHE).

Our work examines the question: \textbf{if a privacy mechanism is already in use, can this be leveraged to provide verification of the LLM inference computation as well?} We answer this question in the affirmative; specifically, we propose two simple but novel protocols that use privacy to obtain differing levels of verification guarantees. We examine the costs and security properties of each of these protocols. Although our protocols have limitations and do not offer identical guarantees to those of ZK, we show that they are robust to many varieties of attacks. Moreover, we demonstrate that our `logit fingerprinting with noise' protocol run with an SMPC method, SIGMA \citep{cryptoeprint:2023/1269}, is $\sim15 \times$ faster than the state-of-the-art ZK method for proof of LLM inference on a single forward pass of Llama-2-7B \citep{touvron2023llama2openfoundation}. We further hope that connecting privacy and verification for LLM inference will spur the creation of improved protocols and further research in this area; to this end, we also open-source an implementation of our Protocol 1 using the CrypTen library at \href{https://github.com/louai-ritual/priveri}{https://github.com/louai-ritual/priveri}.

\begin{figure}
    \centering
    \includegraphics[width=\linewidth]{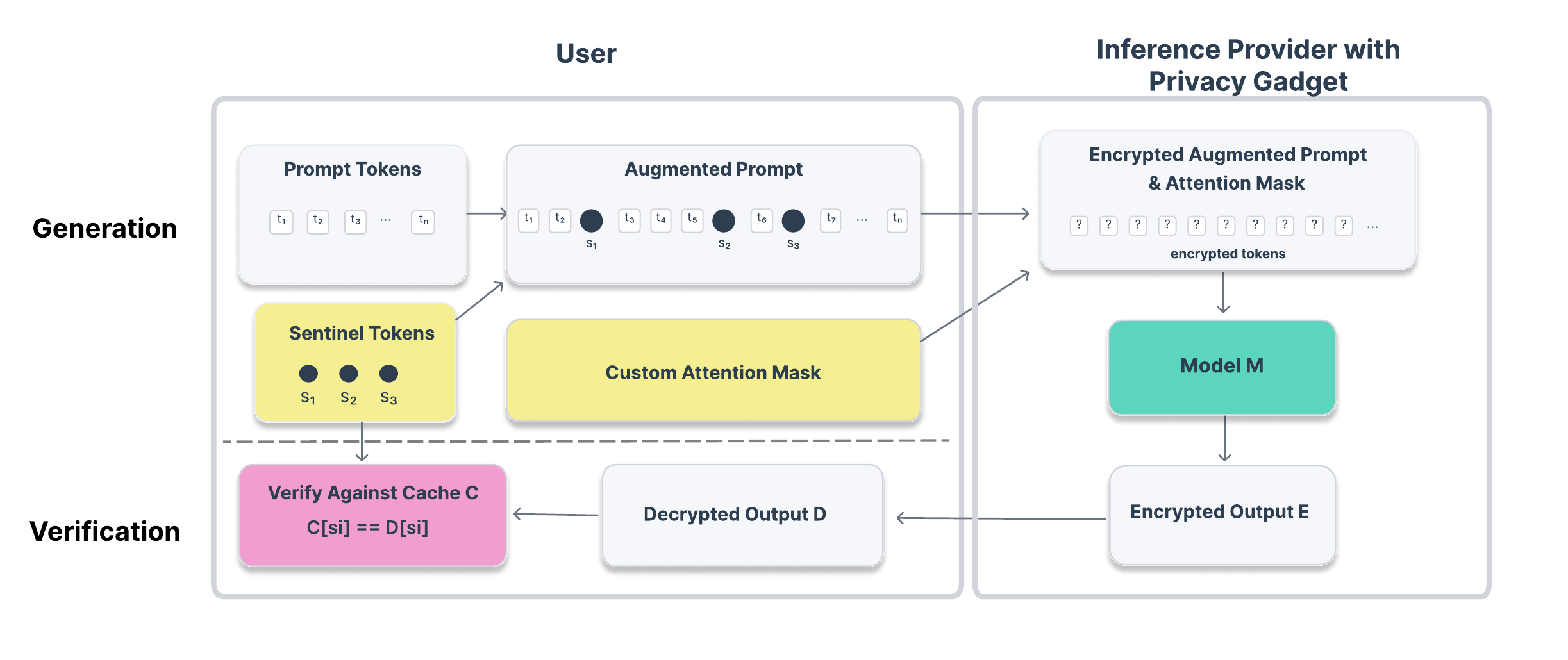}
    \caption{Schema of \textbf{Protocol 1}, which allows a user to verify that a specified model $M$ was run on their prompt by a third party inference provider. The user adds \emph{sentinel tokens} in random positions to the original prompt tokens, along with an altered attention mask. The position of these sentinel tokens is obfuscated to the inference provider by the privacy-preserving mechanism. The user can therefore perform verification against a precomputed cache of the sentinel token logits, effectively ensuring that model substitution attacks cannot occur.}
    \label{fig:figure_1}
\end{figure}

\section{Background \& Related Work}

\textbf{Privacy-preserving inference.} There are four main families of privacy-preserving LLM inference methods that have been proposed in the literature: SMPC (Secure Multi-Party Computation), FHE (Fully Homomorphic Encryption), TEEs (Trusted Execution Environments), and statistical methods. SMPC and FHE are general privacy-preserving computation methods which provide strong guarantees on computational indistinguishability of the inputs. Both methods add significant overhead to plaintext computation; for SMPC, a large component of this is communication between the multiple parties involved. Recently, both SMPC \citep{huang2022cheetah, hao2022iron, pang2023bolt, akimoto2023privformer, dong2023pumasecureinferencellama7b, li2023mpcformerfastperformantprivate} and FHE \citep{moon2024thorfhe, zhang2024nexusfhe} have been applied to LLM inference. Our protocols are agnostic to the exact method used, though differing attacks are possible with each choice of privacy mechanism. For a more detailed account of these methods, see \cref{app:background_privacy}.

\textbf{Verifiable inference.} Zero-knowledge proofs (ZKPs) are a class of methods that allows one party, the prover, to prove to another party, the verifier, that a statement is true, without revealing any additional information beyond the proof itself. ZK methods have recently been applied to proving LLM inference, such as in \citet{sun2024zkllmzeroknowledgeproofs, qu2025zkgpt}. However, these approaches have significant overhead, and remain thousands of times slower than vanilla inference. By contrast, recent work has introduced `statistical' methods for verifiable LLM inference, where the guarantees are relaxed in order to reduce the overhead added. For a more detailed account of these methods, see \cref{app:background_verification}.

\textbf{Connections between privacy and verification.} The connection between privacy and verification has not been extensively studied previously. Perhaps the closest work is MPC-in-the-Head \citep{ishai2007zero}, which introduced a zero-knowledge verification protocol by utilizing any SMPC protocol. The protocol comes with steep costs for both the prover and verifier. For example, the prover must not only locally simulate every party in the underlying MPC execution but also repeat the computation multiple times. On the verifier's end, the party must perform several confirmation tasks, including recomputing opened views, consistency checks, and typically engage in multiple rounds of checking to achieve acceptable soundness. The crucial distinction of our suggested protocols to MPC-in-the-Head is that we use the privacy scheme \emph{directly} to encode inexpensive secrets that are easily verifiable. To the best of our knowledge, there has not previously been any work that specifically examines the relationship between privacy-preserving LLM inference and verifiable inference of LLMs in this way.

\section{Threat Model}
\label{sec:thread_model}
We consider a setting with two primary roles: the \textbf{user}, who also acts as the verifier, and an \textbf{inference provider}, who also acts as the prover. The user wishes to run inference with a model $M$ on their prompts $x$, but cannot do so themselves due to e.g. lack of computational resources. They therefore request the inference provider to perform inference on $x$ with $M$. The inference provider is untrusted and may act as an adversary without behavioral constraints; other external adversaries are out of scope. We assume the use of a privacy-preserving mechanism providing computational indistinguishability of the inputs to ensure that the inference provider cannot view $x$. The model weights are assumed to be public. Our security goal is verifiability -- that is, ensuring that the output the inference provider returns to the user can be verified as being the correct forward pass on the requested model on the given privatized prompt. 




\section{Protocol 1: Logit Fingerprinting}
\label{sec:logit}

Our first proposal for obtaining verification cheaply given access to a privacy-preserving method of LLM inference is \textbf{logit fingerprinting}. We hypothesize that the logit vector returned by performing a forward pass on any set of tokens on modern LLMs is a highly unique `fingerprint' of the model. Our proposed protocol leverages this property to provide inference verification as follows (see \cref{fig:figure_1}):

\begin{enumerate}
    \item First, the user inserts $K$ \emph{sentinel tokens} into the tokenized prompt, at random positions within the prompt. Call these positions $p_1, p_2, ..., p_K$. These $K$ tokens are taken randomly from a public cache $C$, consisting of many such length $K$ sequences.
    \item Next, the user creates the 2D attention mask to be used by the LLM by taking their desired attention mask (e.g., lower triangular for decoder-only LLMs) and inserting rows and columns as follows.
    \begin{itemize}
        \item Add a row at $p_i$ that is 0 everywhere except positions $p_j \, \forall \, j \leq i$, where it is set to 1.
        \item Add a column at $p_i$ that is 0 everywhere except positions $p_j \, \forall \, j \geq i$, where it is set to 1.
    \end{itemize}
    \item The attention mask and augmented tokenized prompt are given to the inference provider under a privacy-preserving scheme, and the inference provider carries out a forward pass, and returns the output logit vector at all token positions to the user.
    \item The user verifies that the sentinel token logits match against the precomputed cached logits for that specific model.
\end{enumerate}

The construction of the attention mask is such that the sentinel tokens do not attend to, and are not attended by, any of the original prompt tokens, but they do attend to each other in standard autoregressive fashion. This also ensures that sentinel tokens have no downstream impact on the original prompt when inference is performed. A formal description of this procedure is given in \cref{app:formal_algo_1}.


\subsection{Cost Analysis}
\label{subsec:logit_cost}

\textbf{Inference provider (prover).} Excluding the overhead of the private inference scheme, the total number of extra operations is a factor of $\frac{K}{N}$, where $N$ is the length of the original prompt. As we discuss in \cref{subsec:logit_security}, $K$ can be set to be as small as 3 and retain strong security properties, so this is very small for reasonably sized $N$. Furthermore, if the privacy scheme supports parallelized inference, this can result in almost no extra runtime.

\textbf{User (verifier).} The verifier is required to pick a sequence from a public cache and perform a matching on the returned logits against the same cache. The cost of this is minimal and does not require specialized hardware. 

\textbf{Construction of the cache.} 
Constructing the cache entails an initial computational cost and also must be performed by a trusted party, since it underpins the correctness of the protocol. Ideally, this responsibility is delegated to an entity with sufficient computational resources to produce a verifiable proof of correctness, for example, in the form of a zero-knowledge proof. Although the computational expense of this might be significant, the cost is incurred only once and is then amortized across all subsequent inference calls, including potentially all prover-verifier pairs.

\subsection{Security Analysis}
\label{subsec:logit_security}

In this section, we assume that logits are indeed unique fingerprints of models. We perform analysis across a range of models in \cref{subsec:logit_experiments} to verify this is the case. 

In order for the inference provider to not be able to guess the logits to return for the sentinel tokens, the set of sentinel tokens must be randomly chosen from a large set of possibilities. The crux of this protocol is that the inference provider cannot determine which of the possibilities is specifically being asked for in any particular instance due to the privacy mechanism.


\textbf{Probabilistic attacks.} This protocol utilizes two elements of randomization: the choice of the sentinel tokens, and their positions. For the former, if the user selects the sequence uniformly at random from a cache of size $|C|$, then a dishonest inference provider can guess it with probability $1/|C|$. $|C|$ can therefore be set to desired tolerances. For the latter, under a privacy-preserving mechanism that also preserves tensor structure (such as SMPC), correctly guessing the sentinel tokens' exact positions is sufficient for a successful attack: the inference provider can perform the forward pass on only those components, and return arbitrary values for the other token positions. However, this occurs with probability $\binom{N + K}{K}^{-1}$, where $N$ is the length of the original prompt. When $K=3$, for example, with $N=14$, this is less than $1\mathrm{e}{-3}$, and it drops further with increasing $N=100$ to circa $1\mathrm{e}{-6}$. 

A related attack is to perform computation only on a random subset of the token indices. In the most extreme case, a dishonest provider takes $N + K - 1$ tokens, i.e. excludes exactly one token. The probability that all sentinel tokens are still selected (hence successfully passing verification) is $\frac{N}{N + K}$, requiring an infeasibly large $K$ to make secure -- although it should be noted that in this case the dishonest provider is saving very little computation over honest behavior.

\textbf{Approximation attacks.} Another line of possible attacks are attempts by the inference provider to use a different model -- especially, cheaper-to-run replacements -- that still succeed in passing verification. Such alternatives could include smaller models from the same model family or approximations to the models by using e.g. low-rank projections of the weights. We perform experiments to test the robustness of the protocol to each of the above in \cref{subsec:logit_experiments} and find that verification fails immediately when any of the above are attempted.

\subsection{Experiments}
\label{subsec:logit_experiments}

\textbf{Setup.} 
We test the claim from \cref{subsec:logit_security} that pre–softmax logits can serve as model fingerprints.
For each model $m$, we sample $N=50{,}000$ token sequences of fixed length $K=3$ from the model’s token vocabulary (excluding special tokens).
Given a sequence $\mathbf{t}=(t_1,t_2,t_3)$, we run a forward pass and record the next-token \emph{logit vectors} at each position, $\ell^{(k)}_m(\mathbf{t}) \in \mathbb{R}^{V_m}$ for $k\!\in\!\{1,2,3\}$, where $V_m$ is the vocabulary size of model $m$.
We define the \emph{logit fingerprint}
\[
\phi_m(\mathbf{t}) \;=\; \operatorname{concat}\!\big(\ell^{(1)}_m(\mathbf{t}),\,\ell^{(2)}_m(\mathbf{t}),\,\ell^{(3)}_m(\mathbf{t})\big) \in \mathbb{R}^{3V_m},
\]
and compare fingerprints using L1 distance. We test on Llama 3.2 Instruct 1B, 3B, and 8B \citep{grattafiori2024llama3herdmodels}, and on Qwen 2.5 Instruct 0.5B, 1.5B, 3B and 7B \citep{qwen2025qwen25technicalreport}. Comparisons are performed on FP32 logits; dropout is disabled.

\begin{table}[t]
\centering
\begin{tabular}{l rr}
\textbf{Model Approximation} & \textbf{Llama} & \textbf{Qwen}\\
\toprule
Same model, GPU non-determinism & 4.08 & 10.91 \\
\midrule
Same model, different sequence & 2909 & 68326    \\
Cross-Model, same sequence & 329096 & 643719\\
Low-rank factorization ($r=2047$) & 833 & 7223 \\
Low-rank factorization ($r=2040$) & 8029 & 124195\\
Low-rank factorization ($r=2000$) & 31194 & 216765 \\
8-bit Quantization & 3499 & 8746 \\
Single step of finetuning & 471 & 1090 \\
\bottomrule
\end{tabular}
\caption{L1 distances of logit fingerprints across different experimental settings for the Llama and Qwen families of models. \textbf{Honest behavior (GPU non-determinism) is clearly separated from dishonest behavior (all other rows).}}
\label{tab:logit_fingerprinting}
\end{table}

\subsubsection{Honest Behaviors}

\textbf{Floating point non-determinism.} We first provide context on the expected L1 distance due to non-determinism of floating-point operations \citep{shanmugavelu2024impactsfloatingpointnonassociativityreproducibility}. This can be constituted as honest behavior; although, it is possible to also require an exact match, which would entail detecting hardware and batched-inference deviations. We run the \emph{same} sequence multiple times with different batch sizes on GPU to measure this. We observe a maximum L1 deviation in doing so across all models tested of $10.90$.

\subsubsection{Dishonest Behaviors}

We now test a wide range of dishonest behaviors and strategies.

\textbf{Intra-model.}
Within each model, we compute the nearest-neighbor similarity among fingerprints from \emph{distinct} sequences (i.e. $\mathbf{t}\neq\mathbf{s}$).
Across $N=50$k samples per model, there are no exact matches; the closest pair has an L1 distance of $2909$.

\textbf{Within-family.} 
For the Llama family, the smallest L1 distance of logits we obtain is $329096$. For the Qwen family, the minimum cross-model distance is $643719$. These results indicate that even with a family of models, the logits are significantly different and suitable as fingerprints.

\textbf{Cross-family.} 
To enable comparisons across families with different vocabularies, we align dimensions by truncating the larger logit vectors to the smaller vocabulary size (i.e. comparing the first $\min(V_m,V_{m'})$ coordinates).
Under this conservative alignment, Llama–Qwen comparisons exhibit substantially higher distances than the within-family maxima reported above (qualitatively, well above $800000$).

\textbf{Low-rank factorization.} 
We approximate the linear layers of Llama~3.2~1B Instruct by replacing each weight matrix $W \in \mathbb{R}^{d_{\text{in}}\times d_{\text{out}}}$ with a rank-$r$ factorization $W \approx U V^\top$, where $U \in \mathbb{R}^{d_{\text{in}}\times r}$ and $V \in \mathbb{R}^{d_{\text{out}}\times r}$.
The default hidden dimension of this model is $2048$, so we test with $r \in \{2047, 2040, 2000\}$.
Comparing fingerprints of 50k sequences between the full-rank and the low-rank variants, the minimum L1 distances observed are:
\[
r=2047:833.97,\quad r=2040:8029.45,\quad r=2000:31194.02.
\]
Similarly, we obtained the following results for Qwen~2.5~3B~Instruct:
\[
r=2047:7223,\quad r=2040:124195,\quad r=2000:216765.
\]

\textbf{Quantization.} 
We next load Llama~3.2~1B Instruct and Qwen~2.5~3B~Instruct in 8-bit precision using \texttt{bitsandbytes} and compare fingerprints to the full-precision (bfloat16) baseline.
The minimum L1 distance is $3499$ for Llama and $8746$ for Qwen, again easily separated from the original model.

\textbf{Fine-tuning.} 
Finally, we evaluate robustness against model fine-tuning by comparing each of Llama~3.2~1B~Instruct and Qwen~2.5~3B~Instruct with a finetuned variant of each corresponding model on a single sample from FineWeb dataset for a single step.
The minimum observed distance is $471$ for Llama and $1090$ for Qwen, consistent with the previous cases and again easily separable from the original model.

Our results are summarized in \cref{tab:logit_fingerprinting}. The minimum L1 distance observed under dishonest behavior is $471$, as seen in Llama's finetuning setting, while the maximum deviation with honest behavior due to floating point non-determinism is only $10.90$ for Qwen. The significant difference allows for clear identification of honest vs. dishonest behavior; based on these results, we recommend using a matching threshold in the range of $15$–$20$. Sequences whose logits differ by less than this threshold can be confidently regarded as originating from the same model; and even a single step of fine-tuning is easily detectable with this threshold.

\subsection{Removing Interaction}

As described above, the protocol requires per-decoding-step interaction between the user and the inference provider. In particular, the user needs to perform three steps for each forward pass: 

\begin{enumerate}
    \item Construction of the (encrypted) augmented prompt and attention mask.
    \item Verification of the logits returned by the inference provider by matching sentinel token positions against the logit cache.
    \item Sampling from the logit vector to obtain the next token.
\end{enumerate}
    
In certain settings, such interaction may be undesirable -- for example, when there are large communication overheads due to low bandwidth or high latency. Therefore we now discuss how to remove interaction from our protocol.

Removing per-step verification is straightforward; the user can simply perform the verification for each forward pass as a batch process once generation has completed. For instance, if 10 tokens are generated, the user verifies the logit vector returned for each of the 10 forward passes, and if any one of the verifications fail, they conclude that the inference provider was dishonest.

To construct the inputs for the inference provider -- the augmented prompt and attention mask -- without interaction, note that the crux of construction's security is that the sentinel token positions $p_1, p_2, ..., p_K$ for the $i$th forward pass must be sampled from $\{1, ..., N_i\}$, where $N_i$ is the length of the prompt at the $i$th forward pass; otherwise, if the positions are sampled from say $N_1$, the inference provider will know that they can output incorrect logits for any token positions that are greater than $N_1$, and they will not be identified. The user can therefore pre-generate positions $p_1, ..., p_K$ sampled from $\{1, ..., N_i\}$ for for $i$ in $\{1, 2, ..., M\}$, where $M$ is the upper bound of tokens generated (for example, the maximum context length of the model). These positions can then be sent to the inference provider as ciphertext, who uses them to build the augmented prompt and attention mask for all generation steps using the same privacy-preserving method. This eliminates the requirement for per-step interaction to obtain new inputs. However, offloading the construction of the inputs to the inference provider adds the potential for dishonesty at this step. Fortunately, the standard logit-cache verification is sufficient to detect any such dishonesty -- if the inputs are not constructed correctly from the pre-generated positions at each step, then the verification of the sentinel tokens will not pass at that step.

The last remaining issue is to ensure that the sampling of the generated token from the logit vector is carried out correctly by the inference provider. This is trivial in the case of 0-temperature sampling -- the user can easily verify post-hoc whether the generated tokens are indeed the argmax of the logit vectors produced at each step. However, for temperatures greater than 0, it is more difficult to ensure that the sampling was performed correctly; as long as there is a non-zero probability of a token being generated, the inference provider can always claim that that was what they sampled, whether they did so correctly or not. One solution is to note that sampling is a small and cheap enough operation such that it is feasible to construct a ZK-proof for each step of generation with minimal overhead. Additionally, we note that a dishonest provider is not particularly incentivized to perform dishonest sampling as the cost savings are very marginal.

\subsection{Limitations}
\label{subsec:logit_limitations}


The main limitation of this protocol is the vulnerability to the subsetting attack mentioned in \cref{subsec:logit_security}. As such, we recommend that this protocol not be used in isolation with privacy mechanisms that retain tensor structure, such as SMPC methods.

\section{Protocol 2: Logit Fingerprinting With Noise}
\label{sec:logit_plus_noise}

The vulnerability of Protocol 1 to a subsetting attack reduces the space of privacy gadgets that it can be used with. Our second proposed protocol is designed to resist this attack. Our modification consists of adding randomly sampled noise to the token embeddings before they are passed into the LLM for the forward pass, and then using a lightweight predictor on the returned final hidden states to predict the noise that was used. Our proposed protocol is as follows:

\begin{enumerate}
    \item First the user samples noise $b \in \mathbb{R}^d_e$, where $d_e$ is the embedding dimension of the model being used for inference, from a discrete set of possibilities $B$.
    \item The user concatenates the noise to the embedding of the original prompt $e \in \mathbb{R}^{N\times d_e}$ in the $d_e$ dimension, to obtain a tensor $b_e \in \mathbb{R}^{N\times 2d_e}$.
    \item The user applies the previously trained \emph{NoiseEmbedder} module on $b_e$ to obtain $e' \in \mathbb{R}^{N\times d_e}$.
    \item The user sends privatized $e'$, augmented with $K$ randomly-positioned sentinel tokens as in Protocol 1, to the inference provider. 
    \item The inference provider performs the forward pass and returns the final hidden states $h \in \mathbb{R}^{(N + K) \times d_h}$, where $d_h$ is the hidden dimension.
    \item The user applies the logit-projection to the hidden states at the sentinel token positions, and checks the validity of these logits against the cache, as in Protocol 1.
    \item The user applies the previously trained prediction module, the \emph{NoisePredictor}, on $h$ at the non-sentinel positions to obtain estimated $\hat{b}$ at each such position. If each obtained $\hat{b}$ matches the sampled $b$ at that position, \emph{and} the sentinel token logit check passes above, then the user can consider the inference to be verified.
\end{enumerate}

In the above procedure, the sentinel tokens are not modified by the sampled noise, and so can be compared against the cache as in Protocol 1. For the remainder of the tokens, the predicted noise is compared to the sampled noise to verify that the forward pass was indeed carried out on each token position. The complete procedure is formally described in \cref{app:formal_algo_2}.

\subsection{Cost Analysis}
\label{subsec:logit_plus_noise_cost}

\textbf{Inference provider (prover).} The cost to the inference provider is the same in this protocol as in Protocol 1.

\textbf{User (verifier).} In addition to the cost associated with the sentinel tokens, the user must now  generate the sampled noise -- which requires little computational cost -- as well as run the NoiseEmbedder and NoisePredictor.

\textbf{Construction of the cache.} This cost remains the same as in Protocol 1.

\textbf{Training of NoiseEmbedder and NoisePredictor.} The NoiseEmbedder and NoisePredictor modules need to be trained for each different LLM in use. This entails an initial computational cost and also must be performed by a trusted party; however, similarly to the cache construction, this is a one-time cost that is then amortized over all subsequent inference calls on that model.

In \cref{subsec:logit_plus_noise_experiments}, we show that the NoiseEmbedder and NoisePredictor can be simple linear projections, so that the additional cost to the user and the training cost can be made low in practice.

\subsection{Security Analysis}
\label{subsec:logit_plus_noise_security}

We inherit the security analysis of Protocol 1 as it pertains to sentinel tokens -- that is, sentinel tokens remain effective markers of the model that was used for the forward pass and are able to detect even very close replacements. For the non-sentinel tokens, the crux of the protocol's security now rests on the predictability of the injected noise.

Let the sample space size be given by $|B|$, and denote the accuracy of the prediction at token position $n$ under honest inference by $\text{acc}_n := P(\hat{b}_n=b_n\ |\ \text{honest})$.

\textbf{Honest provider (completeness).} If the inference provider is honest, the probability that the user incorrectly rejects the returned computation is given by the probability that there is at least one mismatch in the predicted noise: $P(\text{incorrect rejection}) = 1 - \Pi_{n=1}^N \text{acc}_n$.

\textbf{Dishonest provider (soundness).} If the inference provider is dishonest, the probability that the user incorrectly accepts the returned computation is given by the probability that $\hat{b}_n = b_n$ at all token positions $n$. Due to the privacy-preserving mechanism, the provider cannot know which $b_n$ was used, so the probability that $\hat{b}_n = b_n$ for any particular $n$ is upper bounded by $\frac{1}{|B|}$. In particular, the above implies that in a leave-one-out subsetting attack as described in \cref{subsec:logit_security}, the probability of success is at most $\frac{1}{|B|}$.

\subsection{Experiments}
\label{subsec:logit_plus_noise_experiments}

In this section, we describe a performant and lightweight architecture of the NoiseEmbedder and NoisePredictor; and we demonstrate their performance on Llama-3.2-1B, evaluating on the FineWeb-Edu dataset \citep{lozhkov2024fineweb-edu}.

\textbf{NoiseEmbedder architecture.} This module consists of a learned embedding $E \in \mathbb{R}^{|B| \times d_e}$, and a linear layer that is applied to the concatenation of the learned noise embeddings and the original embedding, and produces a single combined embedding as an output. Therefore the linear layer has a weight matrix: $W \in \mathbb{R}^{2d_e \times d_e}$.

\textbf{NoisePredictor architecture.} This module consists of a linear layer that takes the final hidden layer representations from the forward pass of the LLM and outputs unnormalized logits over the sample space $B$. Therefore the linear layer has a weight matrix: $W \in \mathbb{R}^{d_h \times |B|}$.

We fine-tune the NoiseEmbedder and NoisePredictor modules \textbf{whilst keeping the original model weights frozen}. As we are adding noise to the model embeddings, we train to optimize for both the log-likelihood on the dataset, as well as the classification accuracy of the NoisePredictor, using the cross-entropy loss. Our training objective is therefore given by:


\begin{equation}
\begin{aligned}
\mathcal{L}_{\theta, \phi} \;=\; 
\mathbb{E}_{x,y \sim \mathcal{D}} \;
\mathbb{E}_{b \sim B} \Big[
  - \log f \big(y \mid \text{NoiseEmbedder}_{\theta}(x,b) \big) \;+\; \\
  \hspace*{2em}\lambda \, \mathrm{CE}\big(\text{NoisePredictor}_\phi(f(\text{NoiseEmbedder}_\theta(x,b))),\, b \big)
\Big]
\end{aligned}
\label{eq:logit_plus_noise_objective}
\end{equation}

where $x, y$ are the training data, f is the base model, $\theta$ and $\phi$ are the parameters of the NoiseEmbedder and NoisePredictor respectively, and $\lambda$ is a hyperparameter to be tuned. In practice, we find best results applying the same sampled noise to every token in the sequence. This does not impact the security analysis of \cref{subsec:logit_plus_noise_security}. For further training and hyperparameter details, see \cref{app:logit_plus_noise_training}.

\textbf{Results.} Despite using a very lightweight NoiseEmbedder and NoisePredictor, and not modifying the original model weights at all, we find that we are able to achieve $\sim 99 \%$ classification accuracy with $|B| = 100$ without \emph{any} worsening of the log-loss on the given dataset. In particular, the base model's log-prob is $\sim 3.45$, and  we achieve a held-out evaluation set log-prob of $\sim 3.43$ after training the modules, with noise injected.

\subsection{Limitations}

In comparison with Protocol 1, this protocol is resistant to subset attacks, due to the introduction of noise at each token position. However, this protocol adds extra computational burden to the user -- they must now perform additional NoiseEmbedder and NoisePredictor forward passes. Although we have shown that these can be effective even if comprising just a single linear layer each, there may be some cases where even this extra computational requirement cannot be met. Moreover, the user must now also perform projection of the final hidden states to the logits themselves, necessitating another matrix multiplication. There is also now the additional computational requirement of training the modules prior to deployment, in a trust-secured manner. Finally, although we are able to achieve good accuracy rates of $~99\%$ with $|B|=100$, we have neither perfect soundness nor completeness; we hope that future work is capable of improving on the results we present here.

\section{Properties Under Different Privacy Mechanisms}
\label{sec:properties_different_privacy}

So far in the description of our protocols, we have remained largely agnostic to the privacy mechanism in use. In this section, we expand on the specific properties that our protocols possess under each of the main 3 privacy-preserving mechanisms: FHE, TEEs, and SMPC.

\textbf{FHE  } The interaction of our protocols with FHE is the most straightforward. FHE has only a single party performing the encrypted inference. As shown by our analysis above, our protocols are capable of determining if that party is operating honestly or dishonestly.

\textbf{TEE  } The interaction of our protocols with TEEs is multifaceted and complex. TEEs come equipped with both privacy-preservation (via memory encryption of processes restricted to hardware enclaves) as well as a form of verifiability with an attestation mechanism. Attacks on TEEs can degrade guarantees of the confidentiality, the attestation, or both. Our protocols can provide extra safety against attacks on attestation. In particular, the assumption of the trustworthiness of the hardware vendor in signing the enclave measurements can be relaxed. We provide further details in \cref{app:tee_interaction}.

\textbf{SMPC  } SMPC protocols rely on multiple parties performing computations in order to obtain privacy-preserving inference. A typical assumption necessary for SMPC protocols to preserve their guarantees is that participants are `honest-but-curious' i.e. they perform the prescribed computation faithfully, but they may use any means at their disposal in order to try to determine the content of the encrypted messages they receive. Our protocols allow for a relaxation of this (rather unrealistic) behavioral assumption. In particular, our protocols can detect the case where there are dishonest participants, as long as they do not all collude (if they do all collude, then privacy no longer holds in SMPC). Further analysis of this is given in \cref{app:active_security}.

\section{Outperforming State-Of-The-Art ZK Inference}
\label{sec:performance}

We have shown that privacy-preserving mechanisms can enable verified inference. In this section, we describe the performance of privacy-preserving inference and our protocols, compared to the standard approach of zero-knowledge (ZK) proofs of inference. 

The protocols we proposed in \cref{sec:logit} and \cref{sec:logit_plus_noise} are both compatible with FHE privacy schemes. However, state-of-the-art FHE schemes typically have greater overhead than ZK; for example, THOR \citep{moon2024thorfhe} reports approximately 10 minutes for a single forward pass on an input of 128 tokens with BERT-Base (a model with 110M params), with GPU acceleration. By contrast, zkLLM reports just 74s of prover overhead for a forward pass with an input of 2048 tokens on OPT-125M. However, Protocol 2 (\cref{sec:logit_plus_noise}) is designed to resist tensor subset attacks, and is therefore also compatible for use with SMPC schemes. State-of-the-art SMPC schemes operate much faster than FHE. 

We perform a direct comparison of our protocol with SMPC to zkLLM. We evaluate our protocol across varying generation lengths of \{1, 2, 5, 10, 50, 100\} tokens, on a 131 token input prompt, of which 128 are the original prompt tokens and 3 are sentinel tokens required by our construction. We run zkLLM on Llama-2-7B \citep{touvron2023llama2openfoundation} and measure the total prover time for these response lengths on a machine equipped with an A100 GPU. We compare these results to our own measurements of SIGMA \citep{cryptoeprint:2023/1269}, a 2-party SMPC protocol optimized for GPU acceleration, which we evaluate on the same model and hardware. For zkLLM, we measure the runtime for a 128-token prompt.

Our protocol introduces a small, arguably negligible, communication overhead over SIGMA due to the need to transmit the encrypted augmented sequence, augmented attention mask, and augmented position identifiers to the inference provider. The total communication footprint is
\[
b \cdot (L^{2} + 8L + 15),
\]
where $L$ is the sequence length and $b$ is the datatype size in bytes (e.g., 4 bytes for an \texttt{int32}). This expression accounts for the linear-size augmented prompt and position-id vectors and the quadratic-size augmented attention mask.

We evaluate under the same network assumptions as SIGMA. For the LAN setting, we assume a latency of \(\,0.05\,\mathrm{ms}\,\) and a bandwidth of \(9.4\,\mathrm{Gbps}\). For the WAN setting, we assume a latency of \(60\,\mathrm{ms}\) and a bandwidth of \(305\,\mathrm{Mbps}\). These settings ensure a consistent and fair comparison across network environments.


Our results for generation times are shown in \cref{tab:gen_time}. We see that Protocol 2 under SIGMA is approximately $\sim15\times$ faster than zkLLM, across all response lengths; and that there is relatively little degradation in performance when moving from the LAN to the WAN network setting.

\begin{table}[h!]
\centering
\begin{tabular}{c|c|c}
\textbf{Response Length} &
\textbf{Ours with SIGMA (LAN / WAN)} &
\textbf{zkLLM} \\
\hline
\rule{0pt}{2.3ex}
1   & 22.1 / 22.2   & 369.7  \\
2   & 44.4 / 44.4   & 726.5  \\
5   & 118.3 / 118.9 & 1798.1 \\
10  & 227.7 / 236.6 & 3889.3 \\
50  & 1189.0 / 1194.0 & 19550.8 \\
100 & 2260.0 / 2301.0 & 35860.1 \\
\end{tabular}
\caption{Generation time (seconds) for Llama-2-7B with zkLLM vs.\ our Protocol~2 (SIGMA) for different response lengths under LAN and WAN conditions.}
\label{tab:gen_time}
\end{table}

We also provide results for verification times in \cref{tab:verify_time}. Verification with our protocol consists of a lookup into the logit cache. This runs significantly faster than ZK verification.

\begin{table}[h!]
\centering
\begin{tabular}{c|c|c}
\textbf{Response Length} &
\textbf{Ours with SIGMA} &
\textbf{zkLLM} \\
\hline
\rule{0pt}{2.3ex}
1   & 0.17  & 1.24  \\
2   & 0.35  & 2.51  \\
5   & 0.91  & 6.32  \\
10  & 1.87  & 11.80 \\
50  & 3.73  & 59.24 \\
100 & 18.27 & 117.30 \\
\end{tabular}
\caption{Verification time (seconds) for Llama-2-7B with zkLLM vs.\ our Protocol~2 (SIGMA) for different response lengths.}
\label{tab:verify_time}
\end{table}

\textbf{Discussion.} Our protocol as tested in a like-for-like setting is nearly $15\times$ faster than the state-of-the-art ZK method for proof of LLM inference. However, there are two key differences. First, ZK has fewer security assumptions. Although SMPC guarantees strong computational indistinguishability of its inputs in the non-colluding setting, it is vulnerable when all parties involved are dishonest and collude to pool their secret shares. By contrast, ZK is provably secure regardless of prover behavior assumptions. Second, our protocol still relies on statistical results, such as the accuracy of the NoisePredictor module. Therefore, our inference guarantees are not directly comparable to those produced by ZK methods.

Nevertheless, in settings where non-collusion can be ensured or encouraged, and where statistical guarantees are sufficient, our protocol offers a significant speedup over the state-of-the-art for proof of LLM inference.

\section{Conclusion}

We have introduced two protocols for verifying LLM inference, given the use of privacy-preserving mechanisms. These protocols are cheap for both the prover and the verifier and have little to no downstream impact. Future work may focus on mitigating the limitations of our protocols, for example by improving the statistical guarantees.  We believe that connecting privacy and verifiability, particularly in LLM inference, will inspire future work on new and improved protocols.


\bibliography{iclr2026_conference}

@misc{shanmugavelu2024impactsfloatingpointnonassociativityreproducibility,
      title={Impacts of floating-point non-associativity on reproducibility for HPC and deep learning applications}, 
      author={Sanjif Shanmugavelu and Mathieu Taillefumier and Christopher Culver and Oscar Hernandez and Mark Coletti and Ada Sedova},
      year={2024},
      eprint={2408.05148},
      archivePrefix={arXiv},
      primaryClass={cs.DC},
      url={https://arxiv.org/abs/2408.05148}, 
}

@misc{zheng2024permllmprivateinferencelarge,
      title={PermLLM: Private Inference of Large Language Models within 3 Seconds under WAN}, 
      author={Fei Zheng and Chaochao Chen and Zhongxuan Han and Xiaolin Zheng},
      year={2024},
      eprint={2405.18744},
      archivePrefix={arXiv},
      primaryClass={cs.CR},
      url={https://arxiv.org/abs/2405.18744}, 
}

@misc{yuan2024securetransformerinferenceprotocol,
      title={Secure Transformer Inference Protocol}, 
      author={Mu Yuan and Lan Zhang and Xiang-Yang Li},
      year={2024},
      eprint={2312.00025},
      archivePrefix={arXiv},
      primaryClass={cs.CR},
      url={https://arxiv.org/abs/2312.00025}, 
}

@misc{luo2024centaurbridgingimpossibletrinity,
      title={Centaur: Bridging the Impossible Trinity of Privacy, Efficiency, and Performance in Privacy-Preserving Transformer Inference}, 
      author={Jinglong Luo and Guanzhong Chen and Yehong Zhang and Shiyu Liu and Hui Wang and Yue Yu and Xun Zhou and Yuan Qi and Zenglin Xu},
      year={2024},
      eprint={2412.10652},
      archivePrefix={arXiv},
      primaryClass={cs.LG},
      url={https://arxiv.org/abs/2412.10652}, 
}

@INPROCEEDINGS{yao1982protocolssecurecomputations,
  author={Yao, Andrew C.},
  booktitle={23rd Annual Symposium on Foundations of Computer Science (sfcs 1982)}, 
  title={Protocols for secure computations}, 
  year={1982},
  volume={},
  number={},
  pages={160-164},
  doi={10.1109/SFCS.1982.38}}

@inproceedings{goldreich1987howtoplay,
author = {Goldreich, O. and Micali, S. and Wigderson, A.},
title = {How to play ANY mental game},
year = {1987},
isbn = {0897912217},
publisher = {Association for Computing Machinery},
address = {New York, NY, USA},
url = {https://doi.org/10.1145/28395.28420},
doi = {10.1145/28395.28420},
booktitle = {Proceedings of the Nineteenth Annual ACM Symposium on Theory of Computing},
pages = {218–229},
numpages = {12},
location = {New York, New York, USA},
series = {STOC '87}
}

@inproceedings{hao2022iron,
  title     = {Iron: Private Inference on Transformers},
  author    = {Meng Hao and Hongwei Li and Hanxiao Chen and Pengzhi Xing and Guowen Xu and Tianwei Zhang},
  booktitle = {Advances in Neural Information Processing Systems},
  volume    = {35},
  pages     = {15718--15731},
  year      = {2022}
}

@inproceedings{huang2022cheetah,
  title     = {Cheetah: Lean and Fast Secure Two-Party Deep Neural Network Inference},
  author    = {Zhicong Huang and Wen-jie Lu and Cheng Hong and Jiansheng Ding},
  booktitle = {31st USENIX Security Symposium (USENIX Security 22)},
  pages     = {809--826},
  year      = {2022}
}

@misc{pang2023bolt,
      author = {Qi Pang and Jinhao Zhu and Helen Möllering and Wenting Zheng and Thomas Schneider},
      title = {{BOLT}: Privacy-Preserving, Accurate and Efficient Inference for Transformers},
      howpublished = {Cryptology {ePrint} Archive, Paper 2023/1893},
      year = {2023},
      url = {https://eprint.iacr.org/2023/1893}
}

@INPROCEEDINGS{akimoto2023privformer,
  author={Akimoto, Yoshimasa and Fukuchi, Kazuto and Akimoto, Youhei and Sakuma, Jun},
  booktitle={2023 IEEE 8th European Symposium on Security and Privacy (EuroS\&P)}, 
  title={Privformer: Privacy-preserving Transformer with MPC}, 
  year={2023},
  volume={},
  number={},
  pages={392-410},
  doi={10.1109/EuroSP57164.2023.00031}}

@misc{dong2023pumasecureinferencellama7b,
      title={PUMA: Secure Inference of LLaMA-7B in Five Minutes}, 
      author={Ye Dong and Wen-jie Lu and Yancheng Zheng and Haoqi Wu and Derun Zhao and Jin Tan and Zhicong Huang and Cheng Hong and Tao Wei and Wenguang Chen},
      year={2023},
      eprint={2307.12533},
      archivePrefix={arXiv},
      primaryClass={cs.CR},
      url={https://arxiv.org/abs/2307.12533}, 
}

@misc{li2024nimbussecureefficienttwoparty,
      title={Nimbus: Secure and Efficient Two-Party Inference for Transformers}, 
      author={Zhengyi Li and Kang Yang and Jin Tan and Wen-jie Lu and Haoqi Wu and Xiao Wang and Yu Yu and Derun Zhao and Yancheng Zheng and Minyi Guo and Jingwen Leng},
      year={2024},
      eprint={2411.15707},
      archivePrefix={arXiv},
      primaryClass={cs.CR},
      url={https://arxiv.org/abs/2411.15707}, 
}

@misc{bitsandbytes,
  title = {BitsAndBytes: Optimized 8-bit and 4-bit matrix multiplication routines},
  author = {BitsAndBytes},
  year = {2025},
  howpublished = {\url{https://github.com/bitsandbytes-foundation/bitsandbytes}},
  note = {Accessed: 2025-01-28}
}

@misc{li2023mpcformerfastperformantprivate,
      title={MPCFormer: fast, performant and private Transformer inference with MPC}, 
      author={Dacheng Li and Rulin Shao and Hongyi Wang and Han Guo and Eric P. Xing and Hao Zhang},
      year={2023},
      eprint={2211.01452},
      archivePrefix={arXiv},
      primaryClass={cs.LG},
      url={https://arxiv.org/abs/2211.01452}, 
}

@misc{qwen2025qwen25technicalreport,
      title={Qwen2.5 Technical Report}, 
      author={Qwen and : and An Yang and Baosong Yang and Beichen Zhang and Binyuan Hui and Bo Zheng and Bowen Yu and Chengyuan Li and Dayiheng Liu and Fei Huang and Haoran Wei and Huan Lin and Jian Yang and Jianhong Tu and Jianwei Zhang and Jianxin Yang and Jiaxi Yang and Jingren Zhou and Junyang Lin and Kai Dang and Keming Lu and Keqin Bao and Kexin Yang and Le Yu and Mei Li and Mingfeng Xue and Pei Zhang and Qin Zhu and Rui Men and Runji Lin and Tianhao Li and Tianyi Tang and Tingyu Xia and Xingzhang Ren and Xuancheng Ren and Yang Fan and Yang Su and Yichang Zhang and Yu Wan and Yuqiong Liu and Zeyu Cui and Zhenru Zhang and Zihan Qiu},
      year={2025},
      eprint={2412.15115},
      archivePrefix={arXiv},
      primaryClass={cs.CL},
      url={https://arxiv.org/abs/2412.15115}, 
}

@misc{deepseekai2025deepseekr1incentivizingreasoningcapability,
      title={DeepSeek-R1: Incentivizing Reasoning Capability in LLMs via Reinforcement Learning}, 
      author={DeepSeek-AI and Daya Guo and Dejian Yang and Haowei Zhang and Junxiao Song and Ruoyu Zhang and Runxin Xu and Qihao Zhu and Shirong Ma and Peiyi Wang and Xiao Bi and Xiaokang Zhang and Xingkai Yu and Yu Wu and Z. F. Wu and Zhibin Gou and Zhihong Shao and Zhuoshu Li and Ziyi Gao and Aixin Liu and Bing Xue and Bingxuan Wang and Bochao Wu and Bei Feng and Chengda Lu and Chenggang Zhao and Chengqi Deng and Chenyu Zhang and Chong Ruan and Damai Dai and Deli Chen and Dongjie Ji and Erhang Li and Fangyun Lin and Fucong Dai and Fuli Luo and Guangbo Hao and Guanting Chen and Guowei Li and H. Zhang and Han Bao and Hanwei Xu and Haocheng Wang and Honghui Ding and Huajian Xin and Huazuo Gao and Hui Qu and Hui Li and Jianzhong Guo and Jiashi Li and Jiawei Wang and Jingchang Chen and Jingyang Yuan and Junjie Qiu and Junlong Li and J. L. Cai and Jiaqi Ni and Jian Liang and Jin Chen and Kai Dong and Kai Hu and Kaige Gao and Kang Guan and Kexin Huang and Kuai Yu and Lean Wang and Lecong Zhang and Liang Zhao and Litong Wang and Liyue Zhang and Lei Xu and Leyi Xia and Mingchuan Zhang and Minghua Zhang and Minghui Tang and Meng Li and Miaojun Wang and Mingming Li and Ning Tian and Panpan Huang and Peng Zhang and Qiancheng Wang and Qinyu Chen and Qiushi Du and Ruiqi Ge and Ruisong Zhang and Ruizhe Pan and Runji Wang and R. J. Chen and R. L. Jin and Ruyi Chen and Shanghao Lu and Shangyan Zhou and Shanhuang Chen and Shengfeng Ye and Shiyu Wang and Shuiping Yu and Shunfeng Zhou and Shuting Pan and S. S. Li and Shuang Zhou and Shaoqing Wu and Shengfeng Ye and Tao Yun and Tian Pei and Tianyu Sun and T. Wang and Wangding Zeng and Wanjia Zhao and Wen Liu and Wenfeng Liang and Wenjun Gao and Wenqin Yu and Wentao Zhang and W. L. Xiao and Wei An and Xiaodong Liu and Xiaohan Wang and Xiaokang Chen and Xiaotao Nie and Xin Cheng and Xin Liu and Xin Xie and Xingchao Liu and Xinyu Yang and Xinyuan Li and Xuecheng Su and Xuheng Lin and X. Q. Li and Xiangyue Jin and Xiaojin Shen and Xiaosha Chen and Xiaowen Sun and Xiaoxiang Wang and Xinnan Song and Xinyi Zhou and Xianzu Wang and Xinxia Shan and Y. K. Li and Y. Q. Wang and Y. X. Wei and Yang Zhang and Yanhong Xu and Yao Li and Yao Zhao and Yaofeng Sun and Yaohui Wang and Yi Yu and Yichao Zhang and Yifan Shi and Yiliang Xiong and Ying He and Yishi Piao and Yisong Wang and Yixuan Tan and Yiyang Ma and Yiyuan Liu and Yongqiang Guo and Yuan Ou and Yuduan Wang and Yue Gong and Yuheng Zou and Yujia He and Yunfan Xiong and Yuxiang Luo and Yuxiang You and Yuxuan Liu and Yuyang Zhou and Y. X. Zhu and Yanhong Xu and Yanping Huang and Yaohui Li and Yi Zheng and Yuchen Zhu and Yunxian Ma and Ying Tang and Yukun Zha and Yuting Yan and Z. Z. Ren and Zehui Ren and Zhangli Sha and Zhe Fu and Zhean Xu and Zhenda Xie and Zhengyan Zhang and Zhewen Hao and Zhicheng Ma and Zhigang Yan and Zhiyu Wu and Zihui Gu and Zijia Zhu and Zijun Liu and Zilin Li and Ziwei Xie and Ziyang Song and Zizheng Pan and Zhen Huang and Zhipeng Xu and Zhongyu Zhang and Zhen Zhang},
      year={2025},
      eprint={2501.12948},
      archivePrefix={arXiv},
      primaryClass={cs.CL},
      url={https://arxiv.org/abs/2501.12948}, 
}

@misc{grattafiori2024llama3herdmodels,
      title={The Llama 3 Herd of Models}, 
      author={Aaron Grattafiori and Abhimanyu Dubey and Abhinav Jauhri and Abhinav Pandey and Abhishek Kadian and Ahmad Al-Dahle and Aiesha Letman and Akhil Mathur and Alan Schelten and Alex Vaughan and Amy Yang and Angela Fan and Anirudh Goyal and Anthony Hartshorn and Aobo Yang and Archi Mitra and Archie Sravankumar and Artem Korenev and Arthur Hinsvark and Arun Rao and Aston Zhang and Aurelien Rodriguez and Austen Gregerson and Ava Spataru and Baptiste Roziere and Bethany Biron and Binh Tang and Bobbie Chern and Charlotte Caucheteux and Chaya Nayak and Chloe Bi and Chris Marra and Chris McConnell and Christian Keller and Christophe Touret and Chunyang Wu and Corinne Wong and Cristian Canton Ferrer and Cyrus Nikolaidis and Damien Allonsius and Daniel Song and Danielle Pintz and Danny Livshits and Danny Wyatt and David Esiobu and Dhruv Choudhary and Dhruv Mahajan and Diego Garcia-Olano and Diego Perino and Dieuwke Hupkes and Egor Lakomkin and Ehab AlBadawy and Elina Lobanova and Emily Dinan and Eric Michael Smith and Filip Radenovic and Francisco Guzmán and Frank Zhang and Gabriel Synnaeve and Gabrielle Lee and Georgia Lewis Anderson and Govind Thattai and Graeme Nail and Gregoire Mialon and Guan Pang and Guillem Cucurell and Hailey Nguyen and Hannah Korevaar and Hu Xu and Hugo Touvron and Iliyan Zarov and Imanol Arrieta Ibarra and Isabel Kloumann and Ishan Misra and Ivan Evtimov and Jack Zhang and Jade Copet and Jaewon Lee and Jan Geffert and Jana Vranes and Jason Park and Jay Mahadeokar and Jeet Shah and Jelmer van der Linde and Jennifer Billock and Jenny Hong and Jenya Lee and Jeremy Fu and Jianfeng Chi and Jianyu Huang and Jiawen Liu and Jie Wang and Jiecao Yu and Joanna Bitton and Joe Spisak and Jongsoo Park and Joseph Rocca and Joshua Johnstun and Joshua Saxe and Junteng Jia and Kalyan Vasuden Alwala and Karthik Prasad and Kartikeya Upasani and Kate Plawiak and Ke Li and Kenneth Heafield and Kevin Stone and Khalid El-Arini and Krithika Iyer and Kshitiz Malik and Kuenley Chiu and Kunal Bhalla and Kushal Lakhotia and Lauren Rantala-Yeary and Laurens van der Maaten and Lawrence Chen and Liang Tan and Liz Jenkins and Louis Martin and Lovish Madaan and Lubo Malo and Lukas Blecher and Lukas Landzaat and Luke de Oliveira and Madeline Muzzi and Mahesh Pasupuleti and Mannat Singh and Manohar Paluri and Marcin Kardas and Maria Tsimpoukelli and Mathew Oldham and Mathieu Rita and Maya Pavlova and Melanie Kambadur and Mike Lewis and Min Si and Mitesh Kumar Singh and Mona Hassan and Naman Goyal and Narjes Torabi and Nikolay Bashlykov and Nikolay Bogoychev and Niladri Chatterji and Ning Zhang and Olivier Duchenne and Onur Çelebi and Patrick Alrassy and Pengchuan Zhang and Pengwei Li and Petar Vasic and Peter Weng and Prajjwal Bhargava and Pratik Dubal and Praveen Krishnan and Punit Singh Koura and Puxin Xu and Qing He and Qingxiao Dong and Ragavan Srinivasan and Raj Ganapathy and Ramon Calderer and Ricardo Silveira Cabral and Robert Stojnic and Roberta Raileanu and Rohan Maheswari and Rohit Girdhar and Rohit Patel and Romain Sauvestre and Ronnie Polidoro and Roshan Sumbaly and Ross Taylor and Ruan Silva and Rui Hou and Rui Wang and Saghar Hosseini and Sahana Chennabasappa and Sanjay Singh and Sean Bell and Seohyun Sonia Kim and Sergey Edunov and Shaoliang Nie and Sharan Narang and Sharath Raparthy and Sheng Shen and Shengye Wan and Shruti Bhosale and Shun Zhang and Simon Vandenhende and Soumya Batra and Spencer Whitman and Sten Sootla and Stephane Collot and Suchin Gururangan and Sydney Borodinsky and Tamar Herman and Tara Fowler and Tarek Sheasha and Thomas Georgiou and Thomas Scialom and Tobias Speckbacher and Todor Mihaylov and Tong Xiao and Ujjwal Karn and Vedanuj Goswami and Vibhor Gupta and Vignesh Ramanathan and Viktor Kerkez and Vincent Gonguet and Virginie Do and Vish Vogeti and Vítor Albiero and Vladan Petrovic and Weiwei Chu and Wenhan Xiong and Wenyin Fu and Whitney Meers and Xavier Martinet and Xiaodong Wang and Xiaofang Wang and Xiaoqing Ellen Tan and Xide Xia and Xinfeng Xie and Xuchao Jia and Xuewei Wang and Yaelle Goldschlag and Yashesh Gaur and Yasmine Babaei and Yi Wen and Yiwen Song and Yuchen Zhang and Yue Li and Yuning Mao and Zacharie Delpierre Coudert and Zheng Yan and Zhengxing Chen and Zoe Papakipos and Aaditya Singh and Aayushi Srivastava and Abha Jain and Adam Kelsey and Adam Shajnfeld and Adithya Gangidi and Adolfo Victoria and Ahuva Goldstand and Ajay Menon and Ajay Sharma and Alex Boesenberg and Alexei Baevski and Allie Feinstein and Amanda Kallet and Amit Sangani and Amos Teo and Anam Yunus and Andrei Lupu and Andres Alvarado and Andrew Caples and Andrew Gu and Andrew Ho and Andrew Poulton and Andrew Ryan and Ankit Ramchandani and Annie Dong and Annie Franco and Anuj Goyal and Aparajita Saraf and Arkabandhu Chowdhury and Ashley Gabriel and Ashwin Bharambe and Assaf Eisenman and Azadeh Yazdan and Beau James and Ben Maurer and Benjamin Leonhardi and Bernie Huang and Beth Loyd and Beto De Paola and Bhargavi Paranjape and Bing Liu and Bo Wu and Boyu Ni and Braden Hancock and Bram Wasti and Brandon Spence and Brani Stojkovic and Brian Gamido and Britt Montalvo and Carl Parker and Carly Burton and Catalina Mejia and Ce Liu and Changhan Wang and Changkyu Kim and Chao Zhou and Chester Hu and Ching-Hsiang Chu and Chris Cai and Chris Tindal and Christoph Feichtenhofer and Cynthia Gao and Damon Civin and Dana Beaty and Daniel Kreymer and Daniel Li and David Adkins and David Xu and Davide Testuggine and Delia David and Devi Parikh and Diana Liskovich and Didem Foss and Dingkang Wang and Duc Le and Dustin Holland and Edward Dowling and Eissa Jamil and Elaine Montgomery and Eleonora Presani and Emily Hahn and Emily Wood and Eric-Tuan Le and Erik Brinkman and Esteban Arcaute and Evan Dunbar and Evan Smothers and Fei Sun and Felix Kreuk and Feng Tian and Filippos Kokkinos and Firat Ozgenel and Francesco Caggioni and Frank Kanayet and Frank Seide and Gabriela Medina Florez and Gabriella Schwarz and Gada Badeer and Georgia Swee and Gil Halpern and Grant Herman and Grigory Sizov and Guangyi and Zhang and Guna Lakshminarayanan and Hakan Inan and Hamid Shojanazeri and Han Zou and Hannah Wang and Hanwen Zha and Haroun Habeeb and Harrison Rudolph and Helen Suk and Henry Aspegren and Hunter Goldman and Hongyuan Zhan and Ibrahim Damlaj and Igor Molybog and Igor Tufanov and Ilias Leontiadis and Irina-Elena Veliche and Itai Gat and Jake Weissman and James Geboski and James Kohli and Janice Lam and Japhet Asher and Jean-Baptiste Gaya and Jeff Marcus and Jeff Tang and Jennifer Chan and Jenny Zhen and Jeremy Reizenstein and Jeremy Teboul and Jessica Zhong and Jian Jin and Jingyi Yang and Joe Cummings and Jon Carvill and Jon Shepard and Jonathan McPhie and Jonathan Torres and Josh Ginsburg and Junjie Wang and Kai Wu and Kam Hou U and Karan Saxena and Kartikay Khandelwal and Katayoun Zand and Kathy Matosich and Kaushik Veeraraghavan and Kelly Michelena and Keqian Li and Kiran Jagadeesh and Kun Huang and Kunal Chawla and Kyle Huang and Lailin Chen and Lakshya Garg and Lavender A and Leandro Silva and Lee Bell and Lei Zhang and Liangpeng Guo and Licheng Yu and Liron Moshkovich and Luca Wehrstedt and Madian Khabsa and Manav Avalani and Manish Bhatt and Martynas Mankus and Matan Hasson and Matthew Lennie and Matthias Reso and Maxim Groshev and Maxim Naumov and Maya Lathi and Meghan Keneally and Miao Liu and Michael L. Seltzer and Michal Valko and Michelle Restrepo and Mihir Patel and Mik Vyatskov and Mikayel Samvelyan and Mike Clark and Mike Macey and Mike Wang and Miquel Jubert Hermoso and Mo Metanat and Mohammad Rastegari and Munish Bansal and Nandhini Santhanam and Natascha Parks and Natasha White and Navyata Bawa and Nayan Singhal and Nick Egebo and Nicolas Usunier and Nikhil Mehta and Nikolay Pavlovich Laptev and Ning Dong and Norman Cheng and Oleg Chernoguz and Olivia Hart and Omkar Salpekar and Ozlem Kalinli and Parkin Kent and Parth Parekh and Paul Saab and Pavan Balaji and Pedro Rittner and Philip Bontrager and Pierre Roux and Piotr Dollar and Polina Zvyagina and Prashant Ratanchandani and Pritish Yuvraj and Qian Liang and Rachad Alao and Rachel Rodriguez and Rafi Ayub and Raghotham Murthy and Raghu Nayani and Rahul Mitra and Rangaprabhu Parthasarathy and Raymond Li and Rebekkah Hogan and Robin Battey and Rocky Wang and Russ Howes and Ruty Rinott and Sachin Mehta and Sachin Siby and Sai Jayesh Bondu and Samyak Datta and Sara Chugh and Sara Hunt and Sargun Dhillon and Sasha Sidorov and Satadru Pan and Saurabh Mahajan and Saurabh Verma and Seiji Yamamoto and Sharadh Ramaswamy and Shaun Lindsay and Shaun Lindsay and Sheng Feng and Shenghao Lin and Shengxin Cindy Zha and Shishir Patil and Shiva Shankar and Shuqiang Zhang and Shuqiang Zhang and Sinong Wang and Sneha Agarwal and Soji Sajuyigbe and Soumith Chintala and Stephanie Max and Stephen Chen and Steve Kehoe and Steve Satterfield and Sudarshan Govindaprasad and Sumit Gupta and Summer Deng and Sungmin Cho and Sunny Virk and Suraj Subramanian and Sy Choudhury and Sydney Goldman and Tal Remez and Tamar Glaser and Tamara Best and Thilo Koehler and Thomas Robinson and Tianhe Li and Tianjun Zhang and Tim Matthews and Timothy Chou and Tzook Shaked and Varun Vontimitta and Victoria Ajayi and Victoria Montanez and Vijai Mohan and Vinay Satish Kumar and Vishal Mangla and Vlad Ionescu and Vlad Poenaru and Vlad Tiberiu Mihailescu and Vladimir Ivanov and Wei Li and Wenchen Wang and Wenwen Jiang and Wes Bouaziz and Will Constable and Xiaocheng Tang and Xiaojian Wu and Xiaolan Wang and Xilun Wu and Xinbo Gao and Yaniv Kleinman and Yanjun Chen and Ye Hu and Ye Jia and Ye Qi and Yenda Li and Yilin Zhang and Ying Zhang and Yossi Adi and Youngjin Nam and Yu and Wang and Yu Zhao and Yuchen Hao and Yundi Qian and Yunlu Li and Yuzi He and Zach Rait and Zachary DeVito and Zef Rosnbrick and Zhaoduo Wen and Zhenyu Yang and Zhiwei Zhao and Zhiyu Ma},
      year={2024},
      eprint={2407.21783},
      archivePrefix={arXiv},
      primaryClass={cs.AI},
      url={https://arxiv.org/abs/2407.21783}, 
}

@misc{sun2024zkllmzeroknowledgeproofs,
      title={zkLLM: Zero Knowledge Proofs for Large Language Models}, 
      author={Haochen Sun and Jason Li and Hongyang Zhang},
      year={2024},
      eprint={2404.16109},
      archivePrefix={arXiv},
      primaryClass={cs.LG},
      url={https://arxiv.org/abs/2404.16109}, 
}

@misc{kimiteam2025kimik2openagentic,
      title={Kimi K2: Open Agentic Intelligence}, 
      author={Kimi Team and Yifan Bai and Yiping Bao and Guanduo Chen and Jiahao Chen and Ningxin Chen and Ruijue Chen and Yanru Chen and Yuankun Chen and Yutian Chen and Zhuofu Chen and Jialei Cui and Hao Ding and Mengnan Dong and Angang Du and Chenzhuang Du and Dikang Du and Yulun Du and Yu Fan and Yichen Feng and Kelin Fu and Bofei Gao and Hongcheng Gao and Peizhong Gao and Tong Gao and Xinran Gu and Longyu Guan and Haiqing Guo and Jianhang Guo and Hao Hu and Xiaoru Hao and Tianhong He and Weiran He and Wenyang He and Chao Hong and Yangyang Hu and Zhenxing Hu and Weixiao Huang and Zhiqi Huang and Zihao Huang and Tao Jiang and Zhejun Jiang and Xinyi Jin and Yongsheng Kang and Guokun Lai and Cheng Li and Fang Li and Haoyang Li and Ming Li and Wentao Li and Yanhao Li and Yiwei Li and Zhaowei Li and Zheming Li and Hongzhan Lin and Xiaohan Lin and Zongyu Lin and Chengyin Liu and Chenyu Liu and Hongzhang Liu and Jingyuan Liu and Junqi Liu and Liang Liu and Shaowei Liu and T. Y. Liu and Tianwei Liu and Weizhou Liu and Yangyang Liu and Yibo Liu and Yiping Liu and Yue Liu and Zhengying Liu and Enzhe Lu and Lijun Lu and Shengling Ma and Xinyu Ma and Yingwei Ma and Shaoguang Mao and Jie Mei and Xin Men and Yibo Miao and Siyuan Pan and Yebo Peng and Ruoyu Qin and Bowen Qu and Zeyu Shang and Lidong Shi and Shengyuan Shi and Feifan Song and Jianlin Su and Zhengyuan Su and Xinjie Sun and Flood Sung and Heyi Tang and Jiawen Tao and Qifeng Teng and Chensi Wang and Dinglu Wang and Feng Wang and Haiming Wang and Jianzhou Wang and Jiaxing Wang and Jinhong Wang and Shengjie Wang and Shuyi Wang and Yao Wang and Yejie Wang and Yiqin Wang and Yuxin Wang and Yuzhi Wang and Zhaoji Wang and Zhengtao Wang and Zhexu Wang and Chu Wei and Qianqian Wei and Wenhao Wu and Xingzhe Wu and Yuxin Wu and Chenjun Xiao and Xiaotong Xie and Weimin Xiong and Boyu Xu and Jing Xu and Jinjing Xu and L. H. Xu and Lin Xu and Suting Xu and Weixin Xu and Xinran Xu and Yangchuan Xu and Ziyao Xu and Junjie Yan and Yuzi Yan and Xiaofei Yang and Ying Yang and Zhen Yang and Zhilin Yang and Zonghan Yang and Haotian Yao and Xingcheng Yao and Wenjie Ye and Zhuorui Ye and Bohong Yin and Longhui Yu and Enming Yuan and Hongbang Yuan and Mengjie Yuan and Haobing Zhan and Dehao Zhang and Hao Zhang and Wanlu Zhang and Xiaobin Zhang and Yangkun Zhang and Yizhi Zhang and Yongting Zhang and Yu Zhang and Yutao Zhang and Yutong Zhang and Zheng Zhang and Haotian Zhao and Yikai Zhao and Huabin Zheng and Shaojie Zheng and Jianren Zhou and Xinyu Zhou and Zaida Zhou and Zhen Zhu and Weiyu Zhuang and Xinxing Zu},
      year={2025},
      eprint={2507.20534},
      archivePrefix={arXiv},
      primaryClass={cs.LG},
      url={https://arxiv.org/abs/2507.20534}, 
}

@inproceedings{gentryfhe2009,
author = {Gentry, Craig},
title = {Fully homomorphic encryption using ideal lattices},
year = {2009},
isbn = {9781605585062},
publisher = {Association for Computing Machinery},
address = {New York, NY, USA},
url = {https://doi.org/10.1145/1536414.1536440},
doi = {10.1145/1536414.1536440},
booktitle = {Proceedings of the Forty-First Annual ACM Symposium on Theory of Computing},
pages = {169–178},
numpages = {10},
location = {Bethesda, MD, USA},
series = {STOC '09}
}

@inproceedings{sabt2015trustedtee,
  title={Trusted execution environment: What it is, and what it is not},
  author={Sabt, Mohamed and Achemlal, Mohammed and Bouabdallah, Abdelmadjid},
  booktitle={2015 IEEE Trustcom/BigDataSE/Ispa},
  volume={1},
  pages={57--64},
  year={2015},
  organization={IEEE}
}

@article{narra2019privacytee,
  title={Privacy-preserving inference in machine learning services using trusted execution environments},
  author={Narra, Krishna Giri and Lin, Zhifeng and Wang, Yongqin and Balasubramaniam, Keshav and Annavaram, Murali},
  journal={arXiv preprint arXiv:1912.03485},
  year={2019}
}

@inproceedings{ishai2007zero,
  title={Zero-knowledge from secure multiparty computation},
  author={Ishai, Yuval and Kushilevitz, Eyal and Ostrovsky, Rafail and Sahai, Amit},
  booktitle={Proceedings of the thirty-ninth annual ACM symposium on Theory of computing},
  pages={21--30},
  year={2007}
}

@misc{sun2025svipverifiableinferenceopensource,
      title={SVIP: Towards Verifiable Inference of Open-source Large Language Models}, 
      author={Yifan Sun and Yuhang Li and Yue Zhang and Yuchen Jin and Huan Zhang},
      year={2025},
      eprint={2410.22307},
      archivePrefix={arXiv},
      primaryClass={cs.LG},
      url={https://arxiv.org/abs/2410.22307}, 
}

@inproceedings{cheon2017homomorphic,
  title={Homomorphic encryption for arithmetic of approximate numbers},
  author={Cheon, Jung Hee and Kim, Andrey and Kim, Miran and Song, Yongsoo},
  booktitle={International conference on the theory and application of cryptology and information security},
  pages={409--437},
  year={2017},
  organization={Springer}
}

@misc{moon2024thorfhe,
      author = {Jungho Moon and Dongwoo Yoo and Xiaoqian Jiang and Miran Kim},
      title = {{THOR}: Secure Transformer Inference with Homomorphic Encryption},
      howpublished = {Cryptology {ePrint} Archive, Paper 2024/1881},
      year = {2024},
      url = {https://eprint.iacr.org/2024/1881}
}

@misc{zhang2024nexusfhe,
      author = {Jiawen Zhang and Xinpeng Yang and Lipeng He and Kejia Chen and Wen-jie Lu and Yinghao Wang and Xiaoyang Hou and Jian Liu and Kui Ren and Xiaohu Yang},
      title = {Secure Transformer Inference Made Non-interactive},
      howpublished = {Cryptology {ePrint} Archive, Paper 2024/136},
      year = {2024},
      url = {https://eprint.iacr.org/2024/136}
}

@misc{thomas2025cascadetokenshardedprivatellm,
      title={Cascade: Token-Sharded Private LLM Inference}, 
      author={Rahul Thomas and Louai Zahran and Erica Choi and Akilesh Potti and Micah Goldblum and Arka Pal},
      year={2025},
      eprint={2507.05228},
      archivePrefix={arXiv},
      primaryClass={cs.LG},
      url={https://arxiv.org/abs/2507.05228}, 
}

@ARTICLE{jauernig2020teeattacks,
  author={Jauernig, Patrick and Sadeghi, Ahmad-Reza and Stapf, Emmanuel},
  journal={IEEE Security \& Privacy}, 
  title={Trusted Execution Environments: Properties, Applications, and Challenges}, 
  year={2020},
  volume={18},
  number={2},
  pages={56-60},
  doi={10.1109/MSEC.2019.2947124}}

@inproceedings{goldwasser1985zkp,
author = {Goldwasser, S and Micali, S and Rackoff, C},
title = {The knowledge complexity of interactive proof-systems},
year = {1985},
isbn = {0897911512},
publisher = {Association for Computing Machinery},
address = {New York, NY, USA},
url = {https://doi.org/10.1145/22145.22178},
doi = {10.1145/22145.22178},
booktitle = {Proceedings of the Seventeenth Annual ACM Symposium on Theory of Computing},
pages = {291–304},
numpages = {14},
location = {Providence, Rhode Island, USA},
series = {STOC '85}
}

@inproceedings{qu2025zkgpt,
  title={zkGPT: An Efficient Non-interactive Zero-knowledge Proof Framework for LLM Inference},
  author={Qu, Wenjie and Sun, Yijun and Liu, Xuanming and Lu, Tao and Guo, Yanpei and Chen, Kai and Zhang, Jiaheng},
  booktitle={34th USENIX Security Symposium (USENIX Security 25)},
  year={2025}
}

@misc{ong2025toploclocalitysensitivehashing,
      title={TOPLOC: A Locality Sensitive Hashing Scheme for Trustless Verifiable Inference}, 
      author={Jack Min Ong and Matthew Di Ferrante and Aaron Pazdera and Ryan Garner and Sami Jaghouar and Manveer Basra and Max Ryabinin and Johannes Hagemann},
      year={2025},
      eprint={2501.16007},
      archivePrefix={arXiv},
      primaryClass={cs.CR},
      url={https://arxiv.org/abs/2501.16007}, 
}

@misc{lozhkov2024fineweb-edu,
    author       = { Lozhkov, Anton and Ben Allal, Loubna and von Werra, Leandro and Wolf, Thomas },  
    title        = { FineWeb-Edu: the Finest Collection of Educational Content }, 
    year         = 2024,  
    url          = { https://huggingface.co/datasets/HuggingFaceFW/fineweb-edu },  
    doi          = { 10.57967/hf/2497 },
    publisher    = { Hugging Face }
}

@misc{loshchilov2019decoupledweightdecayregularization,
      title={Decoupled Weight Decay Regularization}, 
      author={Ilya Loshchilov and Frank Hutter},
      year={2019},
      eprint={1711.05101},
      archivePrefix={arXiv},
      primaryClass={cs.LG},
      url={https://arxiv.org/abs/1711.05101}, 
}

@misc{cryptoeprint:2023/1269,
      author = {Kanav Gupta and Neha Jawalkar and Ananta Mukherjee and Nishanth Chandran and Divya Gupta and Ashish Panwar and Rahul Sharma},
      title = {{SIGMA}: Secure {GPT} Inference with Function Secret Sharing},
      howpublished = {Cryptology {ePrint} Archive, Paper 2023/1269},
      year = {2023},
      url = {https://eprint.iacr.org/2023/1269}
}

@misc{touvron2023llama2openfoundation,
      title={Llama 2: Open Foundation and Fine-Tuned Chat Models}, 
      author={Hugo Touvron and Louis Martin and Kevin Stone and Peter Albert and Amjad Almahairi and Yasmine Babaei and Nikolay Bashlykov and Soumya Batra and Prajjwal Bhargava and Shruti Bhosale and Dan Bikel and Lukas Blecher and Cristian Canton Ferrer and Moya Chen and Guillem Cucurull and David Esiobu and Jude Fernandes and Jeremy Fu and Wenyin Fu and Brian Fuller and Cynthia Gao and Vedanuj Goswami and Naman Goyal and Anthony Hartshorn and Saghar Hosseini and Rui Hou and Hakan Inan and Marcin Kardas and Viktor Kerkez and Madian Khabsa and Isabel Kloumann and Artem Korenev and Punit Singh Koura and Marie-Anne Lachaux and Thibaut Lavril and Jenya Lee and Diana Liskovich and Yinghai Lu and Yuning Mao and Xavier Martinet and Todor Mihaylov and Pushkar Mishra and Igor Molybog and Yixin Nie and Andrew Poulton and Jeremy Reizenstein and Rashi Rungta and Kalyan Saladi and Alan Schelten and Ruan Silva and Eric Michael Smith and Ranjan Subramanian and Xiaoqing Ellen Tan and Binh Tang and Ross Taylor and Adina Williams and Jian Xiang Kuan and Puxin Xu and Zheng Yan and Iliyan Zarov and Yuchen Zhang and Angela Fan and Melanie Kambadur and Sharan Narang and Aurelien Rodriguez and Robert Stojnic and Sergey Edunov and Thomas Scialom},
      year={2023},
      eprint={2307.09288},
      archivePrefix={arXiv},
      primaryClass={cs.CL},
      url={https://arxiv.org/abs/2307.09288}, 
}
\bibliographystyle{iclr2026_conference}

\clearpage
\appendix
\section{Background and Related Work}
\label{app:background}

In this section, we provide a brief background on general methods of privacy-preserving function computation, general methods of verification, and their application to LLM inference in particular.

\subsection{Privacy-Preservation}
\label{app:background_privacy}

There are four main families of privacy-preserving inference of LLMs that have been proposed in the literature: \textbf{SMPC} (Secure Multi-Party Computation), \textbf{FHE} (Fully Homomorphic Encryption), \textbf{TEEs} (Trusted Execution Environments), and \textbf{statistical methods}. Here we provide brief background on each of these.

\textbf{SMPC } SMPC protocols split the required computation among multiple parties. The key ideas were originally developed in the 1980s \citep{yao1982protocolssecurecomputations, goldreich1987howtoplay} and provide mathematical guarantees that no single party can reconstruct the data on their own.  Recently, the methodologies of SMPC have been applied to LLMs \citep{huang2022cheetah, hao2022iron, pang2023bolt, akimoto2023privformer, dong2023pumasecureinferencellama7b, li2024nimbussecureefficienttwoparty}. A difficulty uniformly faced by these protocols is efficient computation of the many non-linearities present in transformer-based LLMs; most of the works attempt to ameliorate this by using piecewise polynomial approximations which are more well-suited for MPC algorithms. However, this approximation leads to degraded inference results, and remains more expensive than direct computation of the non-linearities. The requirement of multiple parties also engenders significant communication overheads, and the further non-collusion requirement among the parties may be difficult to guarantee. 

\textbf{FHE } FHE protocols require only a single party and make use of cryptographic methods to ensure that the result of the computation on the ciphertext is the same as that performed on the plaintext. The adjective `fully' indicates the capability of performing arbitrary computations, not limited to a particular type or complexity. The first plausible construction of an FHE scheme was described in \citet{gentryfhe2009}; a more modern and widely used incarnation is CKKS \citep{cheon2017homomorphic}. Recently, CKKS has been further optimized and applied to LLM inference \citep{moon2024thorfhe, zhang2024nexusfhe}, but similar issues arise with the non-linearities as SMPC methods. The overheads both for linear and non-linear operations are typically even larger than those in the SMPC setting.

\textbf{TEEs} Trusted Execution Environments (TEEs) \citep{sabt2015trustedtee, narra2019privacytee} create secure and isolated enclaves at the hardware level. This ensures confidentiality via memory encryption -- allowing only the process running in the enclave to read the data. Furthermore, TEEs support integrity via attestation mechanisms. However, a significant concern is the vulnerability to side-channel attacks \citep{jauernig2020teeattacks}. Furthermore, attestation is only provided at boot-time and is not equivalent to an ongoing verification process.  This process typically involves the TEE measuring the code and its environment, signing these measurements cryptographically, and sending a report for external verification. However, this is often a one-time check at the start and does not guarantee the integrity of the TEE throughout its execution. Finally, in cloud environments, attestation can rely on the cloud provider's services, which means users must trust the provider's proprietary attestation process without full transparency. This introduces a level of trust in the cloud provider's integrity, as these attestation services can be opaque "black boxes" that are not open to external audit. Moreover, there may be no independent way to verify the boot measurements provided by the cloud provider's infrastructure. The interaction between our protocol and TEEs is further discussed in \cref{app:tee_interaction}.

\textbf{Statistical Methods } A more broad and diverse grouping than the above is what we term `statistical methods'. These are protocols without the mathematical guarantees of FHE or SMPC approaches, or the hardware-based guarantees of TEEs, but that instead employ statistical or empirical arguments to support the difficult of reversing ciphertext. Some ideas in this domain include the use of permutation-based security \citep{zheng2024permllmprivateinferencelarge, yuan2024securetransformerinferenceprotocol, luo2024centaurbridgingimpossibletrinity} or token-sharding based security \citep{thomas2025cascadetokenshardedprivatellm}. These methods typically trade off the stronger guarantees of the above methods for greatly reduced overheads, sometimes approaching similar speeds to vanilla inference.

\subsection{Verification}
\label{app:background_verification}

\textbf{Zero-Knowledge Proofs (ZKP) } ZKPs are a class of methods that allows one party (the prover) to prove to another party (the verifier) that a statement is true, without revealing any additional information beyond the proof itself. The main properties that ZKPs satisfy are completeness (an honest prover can convince a verifier that they performed the work as stated), soundness (a dishonest prover cannot convince a verifier that they performed the work), and the zero-knowledge property of not revealing any further information than the fact the work was done as stated. The first ZK protocol was introduced in 1985 in \citet{goldwasser1985zkp}. Recently, ZK methods have been applied as proofs of inference for machine learning models, and specifically LLMs, in works such as \citet{sun2024zkllmzeroknowledgeproofs, qu2025zkgpt}. However, these approaches remain thousands of times slower than vanilla inference -- for example, zkLLM takes 15 minutes for generating a proof of a single forward pass for Llama-2-13B, compared to milliseconds for vanilla inference.

\textbf{Statistical Methods } Analogously to statistical methods of privacy-preservation, very recent work has investigated methods of relaxing the standard of proof of work provided in order to reduce computational overhead. \citet{ong2025toploclocalitysensitivehashing} encodes and validates the most salient features of the last hidden state tensor of an LLM using a compact, verifiable proof, which is then recomputed in parallel by the verifier. Although the authors demonstrate how to set up a commitment scheme that has relatively little overhead to the prover, and verification is faster than full recomputation thanks to parallelization, there is still a requirement for the verifier to perform a full LLM forward pass, potentially necessitating specialized hardware. \citet{sun2025svipverifiableinferenceopensource} proposes the use of a `proxy task' based on the last hidden layer features of an LLM that can then be utilized by the user to compare to a label that they would expect based on their original input. The method proposed requires trust assumptions from the platform for generation of the proxy-task feature extractor and labeller networks, as well as secret generation/embedding, and adds the overhead of computation to perform all of the above.

\section{Interaction of Protocols with Secure Multiparty Computation (SMPC)}
\label{app:active_security}

In \cref{sec:properties_different_privacy}, we discussed how our protocols can detect dishonest behavior in the case where not all parties collude, under an SMPC privacy-preserving scheme. Here, we break down the various behavioral cases in detail:

\begin{enumerate}[label=\textbf{Case \arabic*:}]
    \item If there is at least a single honest party participating, that party will be able to detect any dishonest party attempting to deviate from the protocol by trying to evaluate secret shares of the private prompt on an incorrect circuit (i.e., wrong model). However, that honest party may not be incentivized to report this dishonesty; our protocols ensure that this dishonest behavior is detected, even if the honest party does not report the deviation.
    \item If all parties are dishonest and colluding, then the privacy-preservation property of SMPC is broken, and our protocols are no longer effective.
    \item If all parties are dishonest but at least one party is non-colluding, then for most SMPC protocols, privacy-preservation holds. In this setting, our protocols can still detect the dishonest behavior.
\end{enumerate}

In summary, in Case 1, our protocols ensures detection of dishonest behavior regardless of incentive structures; and in Case 3, SMPC by itself is insufficient to ensure the correctness of the computation, and our protocols provide this surety, thereby extending the range of behaviors under which faithful computation can be ensured with SMPC.

\section{Interaction of Protocols with Trusted Execution Environments (TEEs)}
\label{app:tee_interaction}

Trusted Execution Environments (TEEs) provide confidentiality through encrypted and isolated execution, and offer a hardware-backed attestation mechanism intended to certify the integrity of the running code. Because our protocol is designed to compose with general privacy-preserving execution mechanisms, TEEs naturally fit within the class of privacy gadgets through which our approach can achieve verifiability.

However, TEE attestation has well-documented limitations. As discussed in \cref{app:background_privacy}, numerous attacks have been demonstrated across multiple TEE architectures, and additional vulnerabilities continue to emerge. These attacks have different properties w.r.t. whether they break confidentiality, verifiability, or both, and the exact extent to which they do either (e.g. some attacks allow for a limited degree of memory modification, but not sufficiently so to replace the entirety of a model’s loaded weights in a VM). Given the wide variety of attacks, and the very particular details around the adversarial behaviors they enable, we refrain in this paper from making broad statements regarding how our protocols would interact with TEE attack surfaces.

A notable structural assumption of TEE-based attestation is that the hardware vendor acts as the root of trust; the system relies on the vendor not signing forged or incorrect enclave measurements. Our protocol reduces reliance on this assumption: by providing an independent cryptographic verification layer, we relax the trust placed in attestational integrity and enable a lower-trust deployment model in which TEEs serve primarily as a confidentiality mechanism rather than as the sole source of correctness.

Finally, we note that TEEs remain appealing in practice because their runtime overhead is substantially lower than that of fully cryptographic approaches such as SMPC or FHE (typically only 5--10\% for LLM inference). In combination with our protocol, TEEs can therefore support a hybrid design that combines their lightweight confidentiality guarantees with cryptographic verifiability that does not depend exclusively on vendor-rooted trust.

\FloatBarrier

\section{Formal Algorithm for Protocol 1}
\label{app:formal_algo_1}
The procedure is comprised of three components: cache generation through \cref{alg:cache_gen}, inference request through \cref{alg:inf_req}, and the verification stage through \cref{alg:verification}.

\begin{algorithm}
\caption{Cache Generation}
\label{alg:cache_gen}
\begin{algorithmic}[1]
\Require model, cache size $|C| \in \mathbb{N}$, sentinel token count $K \in \mathbb{N}$
\Ensure cache: mapping $s \mapsto \ell_{1:K} \in \mathbb{R}^{K\times V}$
\State cache $\gets \emptyset$
\While{$|cache| < |C|$}
  \State $s_{1:K} \gets$ sample with replacement $K$ tokens from $V$
  \State $m_{1:K,1:K} \gets 0$ \Comment{initialize $K\times K$ attention mask}
  \For{$i = 1..K$}
    \For{$j = 1..i$}
      \State $m_{i,j} \gets 1$
    \EndFor
  \EndFor
  \State $\ell \gets \text{model.forward}(s, m)$ \Comment{$\ell \in \mathbb{R}^{K\times V}$}
  \State cache$[s] \gets \ell$
\EndWhile
\State \Return cache
\end{algorithmic}
\end{algorithm}

\begin{algorithm}
\caption{Inference Request}
\label{alg:inf_req}
\begin{algorithmic}[1]
\Require prompt token embeddings $x_{1:N}$, attention mask $a_{1:N}$, sentinel token sequence $s_{1:K}$
\Ensure logits $\ell_{1:N+K}\in \mathbb{R}^{(N+K) \times V}$
\State positions $p_{1:K}$ $\gets$ sample without replacement $K$ times from Uniform$[1, N + K]$
\State augmented embeddings $x'_{1:N+K}$ $\gets$ insert sentinel tokens $s_{1:K}$ at positions $p_{1:K}$
\State augmented mask $a'_{1:N+K}$ $\gets$ expand $a_{1:N}$ at positions $p_{1:K}$ with 0-filled rows and columns
\For{i = 1..K}
\For{j = 1..i}
  \State $a'[p_i, p_j]$ $\gets$ 1
\EndFor
\For{j=i..K}
  \State $a'[p_j, p_i]$ $\gets$ 1
\EndFor
\EndFor
\State $x' \gets$ encrypt($x'$)
\State $a' \gets$ encrypt($a'$)
\State encrypted $\ell_{1:N+K}$ $\gets$ inference provider forward pass on encrypted $x', a'$
\State \Return decrypt($\ell_{1:N+K}$)
\end{algorithmic}
\end{algorithm}

\begin{algorithm}
\caption{Verification}
\label{alg:verification}
\begin{algorithmic}[1]
\Require logits $\ell_{1:N+K} \in \mathbb{R}^{(N+K)\times V}$, sentinel positions $p_{1:K}\subset\{1,2,..N+K\}$, sentinel sequence $s_{1:K}$, $\text{cache} \in \mathbb{R}^{C \times K \times V}$, tolerance $tol>0$
\Ensure verified: bool
\State verified $\gets$ true
\For{$i = 1..K$}
  \State $p' \gets p[i]$
  \State $err \gets \|\ell[p'] - \text{cache}[s][i]\|_1$
  \If{$err > tol$}
    \State verified $\gets$ false
  \EndIf
\EndFor
\State \Return verified
\end{algorithmic}
\end{algorithm}

\FloatBarrier

\section{Formal Algorithm for Protocol 2}
\label{app:formal_algo_2}
The procedure is also comprised of three algorithms similar to those in \cref{app:formal_algo_1}: noised embedding generation \cref{alg:noised_emb_gen}, noisy inference request \cref{alg:noisy_inf_req}, and verification with noisy prediction \cref{alg:noised_verification}.

\begin{algorithm}
\caption{Noised Embedding Generation}
\label{alg:noised_emb_gen}
\begin{algorithmic}[1]
\Require discrete noise set $B$, trained NoiseEmbedder $E$, embedding dim $d_e$, token embedding $e \in R^{d_e}$
\Ensure sampled noise $b \in R^{d_e}$, noised embedding $e_b \in \mathbb{R}^{d_e}$
\State $b \gets$ sample one value uniformly from $B$
\State $b_e \gets \text{concat}(e, b)$ \Comment{$b_e \in \mathbb{R}^{2d_e}$}
\State $e' \gets E.\text{forward}(b_e)$  \Comment{$e' \in \mathbb{R}^{d_e}$}
\State \Return $(b, e')$
\end{algorithmic}
\end{algorithm}

\begin{algorithm}
\caption{Noisy Inference Request}
\label{alg:noisy_inf_req}
\begin{algorithmic}[1]
\Require prompt token embeddings $x_{1:N}$, attention mask $a_{1:N}$, sentinel token sequence $s_{1:K}$
\Ensure hidden states $h \in\mathbb{R}^{(N+K)\times d_h}$, noise\_cache $\in B^{N}$

\State positions $p_{1:K}$ $\gets$ sample without replacement $K$ times from Uniform$[1, N + K]$
\State augmented embeddings $x'_{1:N+K}$ $\gets$ insert sentinel tokens $s_{1:K}$ at positions $p_{1:K}$
\State augmented mask $a'_{1:N+K}$ $\gets$ expand $a_{1:N}$ at positions $p_{1:K}$ with 0-filled rows and columns
\For{i = 1..K}
\For{j = 1..i}
  \State $a'[p_i, p_j]$ $\gets$ 1
\EndFor
\For{j=i..K}
  \State $a'[p_j, p_i]$ $\gets$ 1
\EndFor
\EndFor

\For{each non-sentinel token $t$ in $x'$}
  \State $b, e' \gets$ call \cref{alg:noised_emb_gen} on $x'[t]$
  \State $x'[t] \gets e'$
  \State noise\_cache[t] $\gets$ b
\EndFor

\State $x' \gets$ encrypt($x'$)
\State $a' \gets$ encrypt($a'$)
\State encrypted $h_{1:N+K}$ $\gets$ inference provider forward pass on encrypted $x', a'$
\State $h \gets decrypt(h_{1:N+K}$)
\State \Return $h$, noise\_cache

\end{algorithmic}
\end{algorithm}

\begin{algorithm}
\caption{Verification With Noise Prediction}
\label{alg:noised_verification}
\begin{algorithmic}[1]
\Require decrypted hidden states $h_{1:N+K}\in\mathbb{R}^{(N+K)\times d_h}$, sentinel positions $p_{1:K}$, sentinel sequence $s_{1:K}$, logit\_cache $\in R^{C \times K \times V}$, noise\_cache $\in B^{N}$, NoisePredictor $NP$, logit projection $L:\mathbb{R}^{d_h}\to\mathbb{R}^V$, sentinel tolerance $tol_s$
\Ensure verified: bool
\For{$i = 1..K$}
  \State $p' \gets p_i$
  \State $\ell \gets L(h[p'])$
  \State $err \gets \|\ell - \text{logit\_cache}[s][i]\|_1$
  \If{$err > tol_s$}
    \State verified $\gets$ false
  \EndIf
\EndFor

\For{$j = 1..N+K$}
  \If{$j \in P$} \State \textbf{continue} \EndIf
  \State $\hat b \gets NP.\text{forward}(h[j])$
  \If{$\hat b \neq \text{noise\_cache}[j]$}
    \State verified $\gets$ false
  \EndIf
\EndFor

\State \Return verified
\end{algorithmic}
\end{algorithm}

\FloatBarrier

\section{Training Details for Protocol 2}
\label{app:logit_plus_noise_training}

In this section we provide further details for the experiments conducted in \cref{subsec:logit_plus_noise_experiments}.

We train on the FineWeb-Edu dataset \citep{lozhkov2024fineweb-edu}. This is a large-scale dataset of 1.3T total tokens consisting of high-quality educational web pages filtered from the larger FineWeb dataset. This dataset has been used for pretraining, and is suitable for general testing of language modeling capabilities of LLMs. We take 40000 samples from this dataset and divide these into an 80/20 train-validation split. We perform training for 500 steps with a batch size of 64 on sequences of length 256. We use the optimizer AdamW \citep{loshchilov2019decoupledweightdecayregularization} with a learning rate of $5e-4$, with no warmup steps. The base model is Llama-3.2-1B, and the weights of this model are frozen; gradients are backpropagated through this model in order to reach the NoiseEmbedder module.

We utilize the same sampled noise at every token position, but a different noise is sampled for each batch element. Our results are reported using a $\lambda$ hyperparameter value of 3.5. We train using an A100 GPU.

\end{document}